\documentclass[epj,twocolumn]{webofc}
\usepackage[varg]{txfonts}   
\woctitle{Hadron Collider Physics symposium 2012}
\begin{document}
\title{Latest Constraints from Jet Measurements on Parton Distribution Functions and on the Strong Coupling Constant}

\author{Sebastian Naumann-Emme\inst{1}\fnsep\thanks{\email{sebastian.naumann@desy.de}} (on behalf of the ATLAS, CDF, CMS, D0, H1 and ZEUS Collaborations)}

\institute{Deutsches Elektronen-Synchrotron (DESY), Hamburg, Germany}

\abstract{%
  Recent measurements of jet cross sections at HERA, the Tevatron and the LHC that provide
  constraints on parton distribution functions and allow for determinations of
  the strong coupling constant are presented.
}
\maketitle
\section{Introduction}
\label{sec:intro}
The factorization theorem of Quantum Chromodynamics (QCD) allows to separate cross sections, $\sigma$,
into a hard-scattering matrix element, $\hat \sigma$, on one hand and parton distribution functions (PDFs) on the other.
While $\hat \sigma$, describing the short-distance structure of the reaction, is process-dependent but
perturbatively calculable, the PDFs, which account for the non-perturbative long-distance structure, are universal
but have to be determined from experimental data.

Beside the quark masses, the only free parameter of the QCD Lagrangian is the strong coupling constant, $\alpha_S$.
The renormalization group equation predicts the energy dependence of the strong coupling, i.e. a functional form for
$\alpha_S (Q)$, where $Q$ is the energy transferred in the reaction.
However, the absolute strength remains a free parameter of the theory,
which for convenience is usually compared at the pole mass of the Z boson.
The latest world average is $\alpha_{S}(M_{Z})$ = 0.1184 $\pm$ 0.0007 \cite{PhysRevD.86.010001}.
This average and its remarkable precision are driven by results obtained at relatively low
energies, namely from hadronic decays of $\tau$ leptons and from lattice QCD.

Jet and multijet production rates are directly sensitive to $\alpha_S$,
in principle even allow for measurements of the ``running'' at high energies,
and they provide constraints on PDFs, especially at medium and high parton momentum fractions, $x$.
For both PDF and $\alpha_S$ analyses, the measured cross sections have to be compared to predictions,
which for most jet observables are currently available to next-to-leading order (NLO) in perturbative QCD.

In this article, the latest jet measurements at HERA, the Tevatron and the LHC
with sensitivity to PDFs and $\alpha_S$ are summarized.

\section{Measurements at HERA}
\label{sec:HERA}

\subsection{Deep-inelastic scattering}
\label{sec:DIS}
\begin{figure}
\centering
\includegraphics[width=\columnwidth,viewport=15 25 515 505,clip]{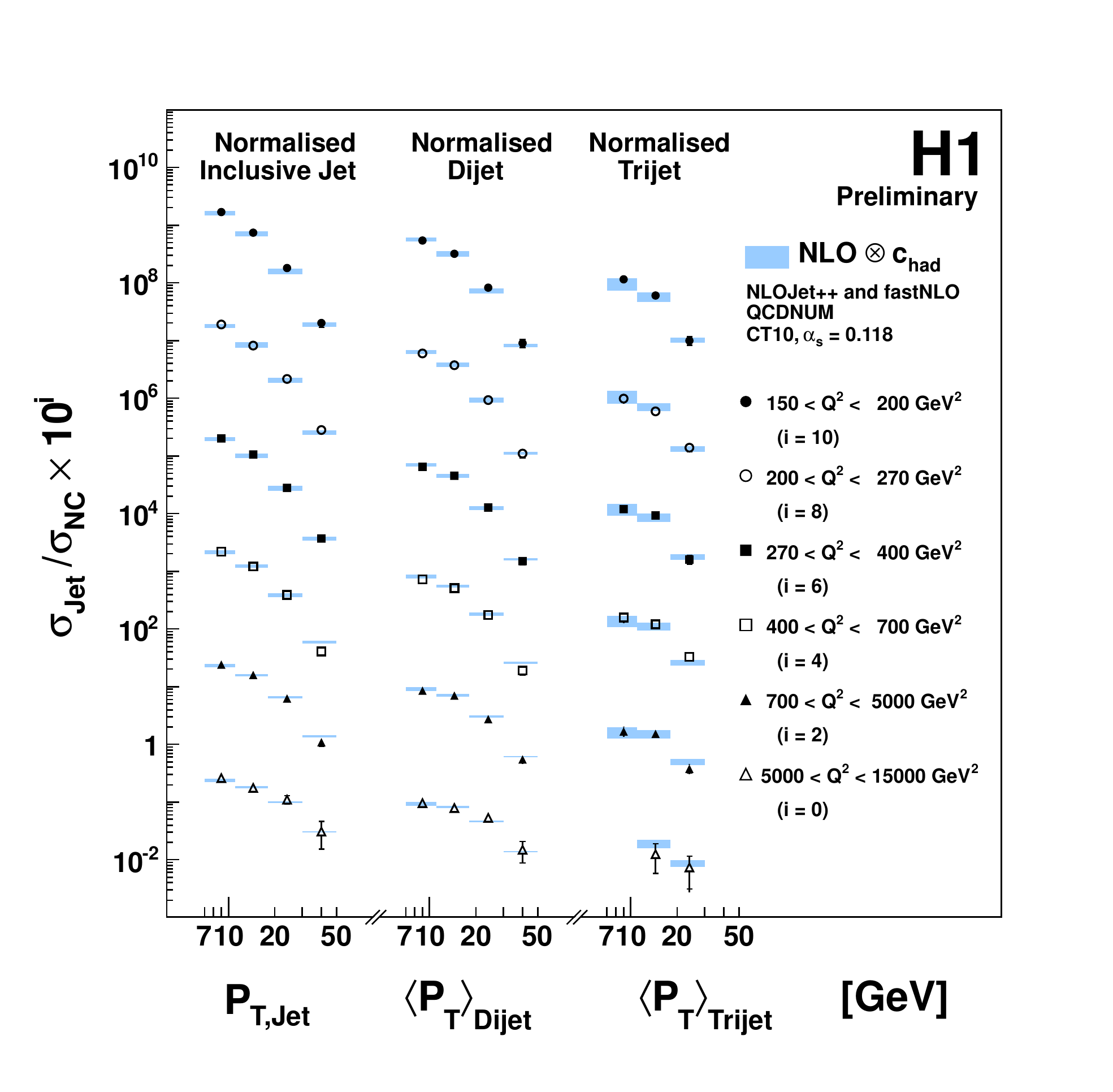}
\caption{Normalized inclusive jet, dijet and trijet cross sections measured by the H1 Collaboration
  as a function of the transverse momentum in different $Q^2$ bins and compared to the NLO prediction
  \cite{H1prelim-12-031}.}
\label{fig:H1prelim-12-031_fig7}
\end{figure}

Based on an earlier publication with double-differential cross sections for
inclusive jet, dijet and trijet production in deep-inelastic scattering (DIS) of electrons/positrons on protons
at a center-of-mass energy, $\sqrt{s}$, of 318~GeV in $Q^2$ bins from 150 to 15000 GeV${}^2$ \cite{H1prelim-11-032},
which confirmed a reasonable description of the data by the NLO prediction and sensitivity to differences in the
proton PDFs,
the H1 Collaboration has recently released updated results based on a regularized unfolding that is performed for all
bins and multijet categories simultaneously \cite{H1prelim-12-031}.
This allows for an optimized correction of detector effects and yields a complete correlation matrix,
valuable for the inclusion of these data into different QCD fits.
The H1 Collaboration used the covariance matrix to normalize the different
jet rates to the inclusive neutral-current DIS cross section. This normalization reduces both
the experimental and the theoretical uncertainties.
The results are shown in figure~\ref{fig:H1prelim-12-031_fig7}.
The cross-section predictions at next-to-leading and leading order were required to agree within 30\%,
which was fulfilled for 42 out of 65 bins of this measurement, before performing a combined fit of $\alpha_S (M_{Z})$
to the normalized inclusive jet, dijet and trijet data.
The result is
$\alpha_S (M_Z)$ = 0.1163 $\pm$ 0.0011 (exp.) $\pm$ 0.0014 (PDF) $\pm$ 0.0008 (hadr.) $\pm$ 0.0039 (th.).

The H1 and ZEUS Collaborations explicitly studied $\alpha_S$-PDF correlations and the impact of HERA jet data on their
PDF fits \cite{H1prelim-11-034}.
When using inclusive DIS data only, taking $\alpha_S$ as a free parameter of the PDF fit increases the uncertainty on
the gluon density at low $x$.
Adding jet data, reduces the correlation between $\alpha_S$ and the gluon PDF and allows for an in-situ $\alpha_S$
determination.
The result obtained within the HERAPDF1.6 fit is $\alpha_S (M_Z)$ = 0.1202 $\pm$ 0.0013 (exp.) $\pm$ 0.0007 (model/param.) 
$\pm$ 0.0012 (hadr.) ${}^{+0.0045}_{-0.0036}$ (scale).

\subsection{Photoproduction}
\label{sec:gammap}
\begin{figure}
\centering
\includegraphics[width=.96\columnwidth,clip]{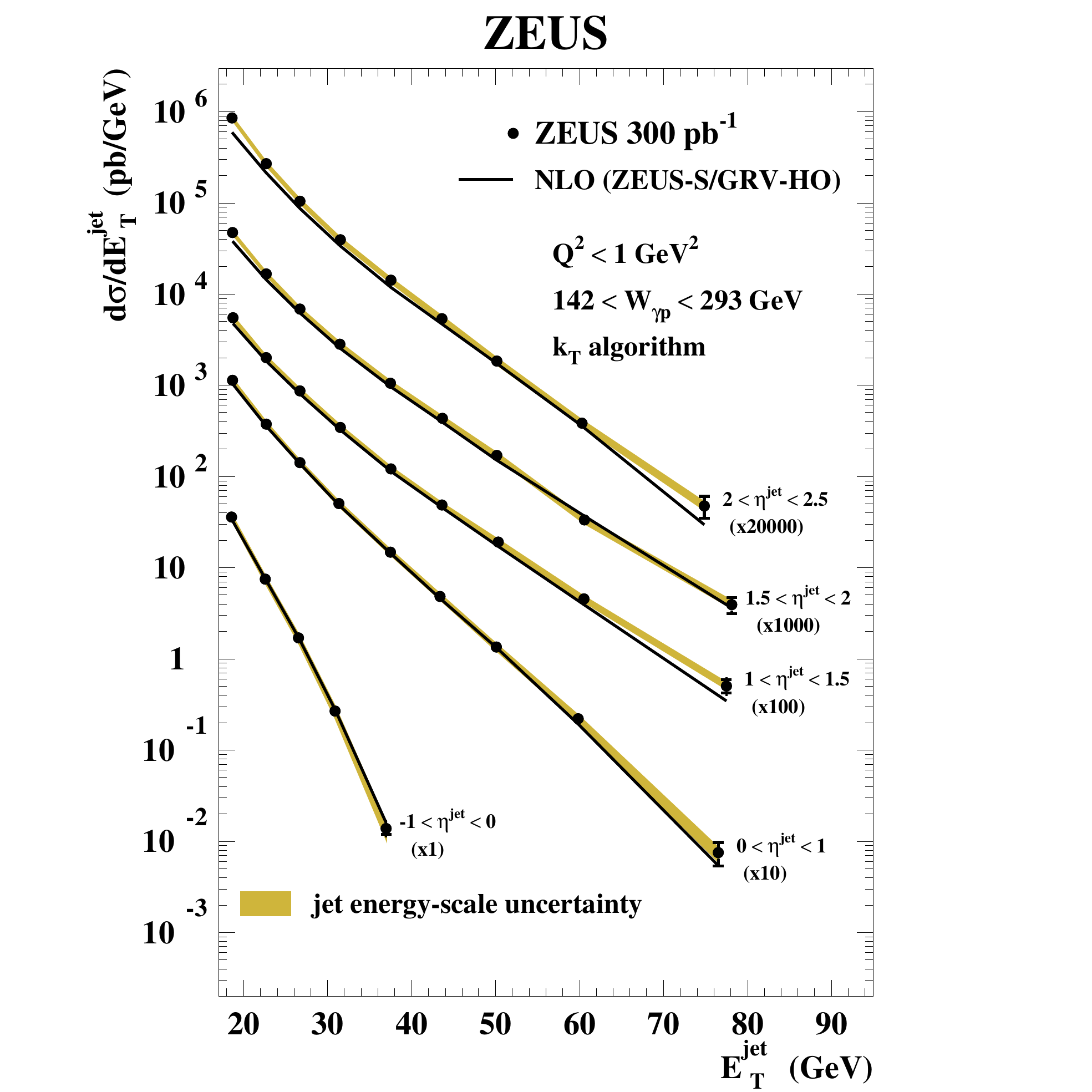}
\caption{Inclusive jet cross section in photoproduction measured by the ZEUS Collaboration as a function of
  the transverse jet energy in different rapidity regions and compared to the NLO prediction \cite{Abramowicz:2012jz}.}
\label{fig:DESY-12-045}
\end{figure}

Besides the DIS regime ($Q^2 >$ 1 GeV${}^2$), where a virtual boson is exchanged,
there is photoproduction ($Q^2 \approx 0$~GeV${}^2$), where a quasi-real photon is emitted from the incoming lepton.
In direct photoproduction, the photon directly takes part in the interaction, while
the photon acts as a source of partons in events with resolved photoproduction.
The latter are rather similar to hadron-hadron collisions since the cross section involves a photon PDF,
there will be a photon remnant and the event can be affected by multi-parton interactions.
As shown in figure~\ref{fig:DESY-12-045},
the ZEUS Collaboration published new double-differential jet cross sections in photoproduction events
with a center-of-mass energy of the photon-proton system between 142 and 293~GeV \cite{Abramowicz:2012jz}.
Compared to DIS measurements, this analysis has a relatively high reach of transverse jet energies, $E_T$, up to 80~GeV.
Overall, a reasonable agreement between data and the NLO prediction is observed.
Discrepancies at low $E_T$ and large pseudorapidities, $\eta$, could come from non-perturbative effects,
in particular related to multi-parton interaction, or the photon PDF.
The impact of the proton PDF uncertainty is in turn much smaller and relevant only at high $E_T$.
The ZEUS Collaboration employed these jet cross sections to measure $\alpha_S$.
The considered range in $d \sigma / d E_T$ was restricted to 21 $< E_T <$ 71~GeV in order to reduce
potential non-perturbative effects and the impact of the proton PDF uncertainty.
Based on the $k_T$ jet algorithm, an $\alpha_S (M_Z)$ of $0.1206\ ^{+0.0023}_{-0.0022}\ {\rm (exp.)}\ ^{+0.0042}_{-0.0035}\ {\rm (th.)}$
was obtained.
The dominant uncertainties are related to the photon PDF and terms beyond NLO.
Results based on the anti-$k_T$ and SIScone algorithms are almost identical.

\section{Measurements at the Tevatron}
\label{sec:Tevatron}
The inclusive jet cross-section in collisions of protons with anti-protons at $\sqrt{s}$ = 1.96~TeV
measured with high precision by the CDF and D0 Collaborations up to jet $p_T$ of 600~GeV
\cite{Aaltonen:2008eq,Abazov:2008ae}, serve as one of the main constraints on the gluon density for $x >$ 0.2-0.3
in global PDF fits already since several years.
Both Collaborations also measured differential dijet and trijet cross sections with significant sensitivity to
differences between the available PDF sets (see e.g. \cite{Abazov:2011ub}).
The inclusion of such multijet cross sections in PDF fits could again help to decorrelate the value of
$\alpha_S$ and the gluon density.

The D0 Collaboration performed an $\alpha_S$ measurement from the inclusive jet cross section \cite{Abazov:2009nc}.
Since the same data had already been used as input for the MSTW2008 PDF set, on which this $\alpha_S$ analysis was based,
the phase space with $x \gtrsim$ 0.25, where the impact on the gluon PDF had been significant, was neglected.
Although that restricted the analysis to 22 data points up to jet $p_T$ of 145~GeV,
the extracted $\alpha_S (M_Z)$ of 0.1161 ${}^{+0.0034}_{-0.0033}$ (exp.) ${}^{+0.0010}_{-0.0016}$ (non-pert.) ${}^{+0.0011}_{-0.0012}$ (PDF)
${}^{+0.0025}_{-0.0029}$ (scale) is still the most precise based on jet data from hadron-hadron collisions to date.

\begin{figure}
\centering
\includegraphics[width=\columnwidth,clip]{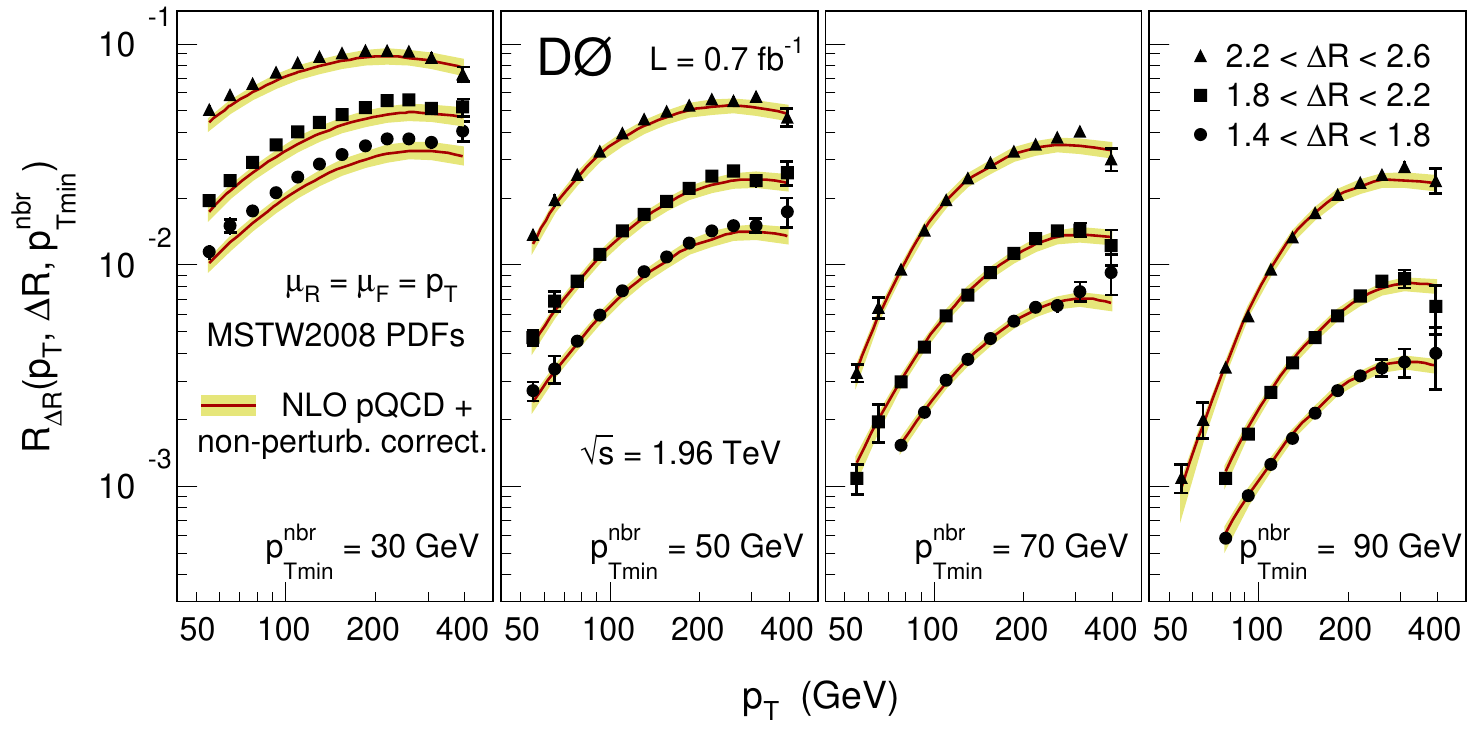}
\caption{Average number of neighboring jets measured by the D0 Collaboration, shown in comparison to the NLO prediction
  and as a function of the inclusive jet transverse momentum, for three different intervals in the jet-jet distance
  and four different requirements of minimum transverse momentum of the neighboring jet \cite{:2012xib}.}
\label{fig:1207-4957}
\end{figure}

Recently, the DO Collaboration performed a new $\alpha_S$ determination using the observable
$R_{\Delta R} (p_T,\Delta R,p_{T \rm min}^{\rm nbr})
= \sum_{i=1}^{N_{\rm jet}(p_T)} N_{\rm nbr}^{(i)}(\Delta R,p_{T \rm min}^{\rm nbr}) / N_{\rm jet}(p_T)$ with $N_{\rm nbr}$ being the number
of neighboring jets within a distance of $\Delta R$ that have a transverse momentum above $p_{T \rm min}^{\rm nbr}$
\cite{:2012xib}.
This normalized triple-differential cross section, shown in figure~\ref{fig:1207-4957},
was found to be well described by the NLO prediction
for $p_{T \rm min}^{\rm nbr}$ = 50~GeV and higher.
$R_{\Delta R}$ allowed probing $\alpha_S (p_T)$ up to jet $p_T$ of 400~GeV
and mostly independent of assumptions on the running.
By combining all data points with $p_{T \rm min}^{\rm nbr}$ = 50, 70 and 90 GeV, an $\alpha_S (M_Z)$ of
0.1191 ${}^{+0.0008}_{-0.0009}$ (exp.) ${}^{+0.0002}_{-0.0001}$ (non-pert.) ${}^{+0.0010}_{-0.0024}$ (PDF)
${}^{+0.0046}_{-0.0066}$ (scale) was obtained.
This result has a remarkable experimental precision and reduced uncertainties related to non-perturbative
corrections but a stronger dependence on the choice of the renormalization and factorization scales,
which emphasizes the need for predictions beyond NLO for multijet topologies.

\section{Measurements at the LHC}
\label{sec:LHC}
Both the ATLAS and the CMS Collaboration measured double-differential cross sections for
inclusive jet and dijet production at $\sqrt{s} =$~7~TeV using up to 37~pb${}^{-1}$ from
the 2010 dataset \cite{Aad:2011fc,CMS:2011ab},
where the lower event pile-up allowed to measure down to relatively low jet momenta and dijet masses,
and using the 5~fb${}^{-1}$ of the
2011 dataset \cite{ATLAS-CONF-2012-021,:2012bz}, where the jet trigger and selection thresholds had
to be raised but the huge amount of data collected yields small statistical uncertainties even at jet momenta up to
2~TeV and dijet masses up to 5~TeV.
Overall, a reasonable agreement is observed between data and the NLO prediction.
Experimental and theoretical uncertainties on these cross sections are of comparable size, where
the jet energy scale is the dominant experimental uncertainty
and PDF uncertainties give a significant contribution to the total uncertainty on the predicted cross sections.
There are some discrepancies between the central predictions obtained with different PDF sets.
However, these differences are well covered by the uncertainties.

\begin{figure}
\centering
\includegraphics[width=.9\columnwidth,clip]{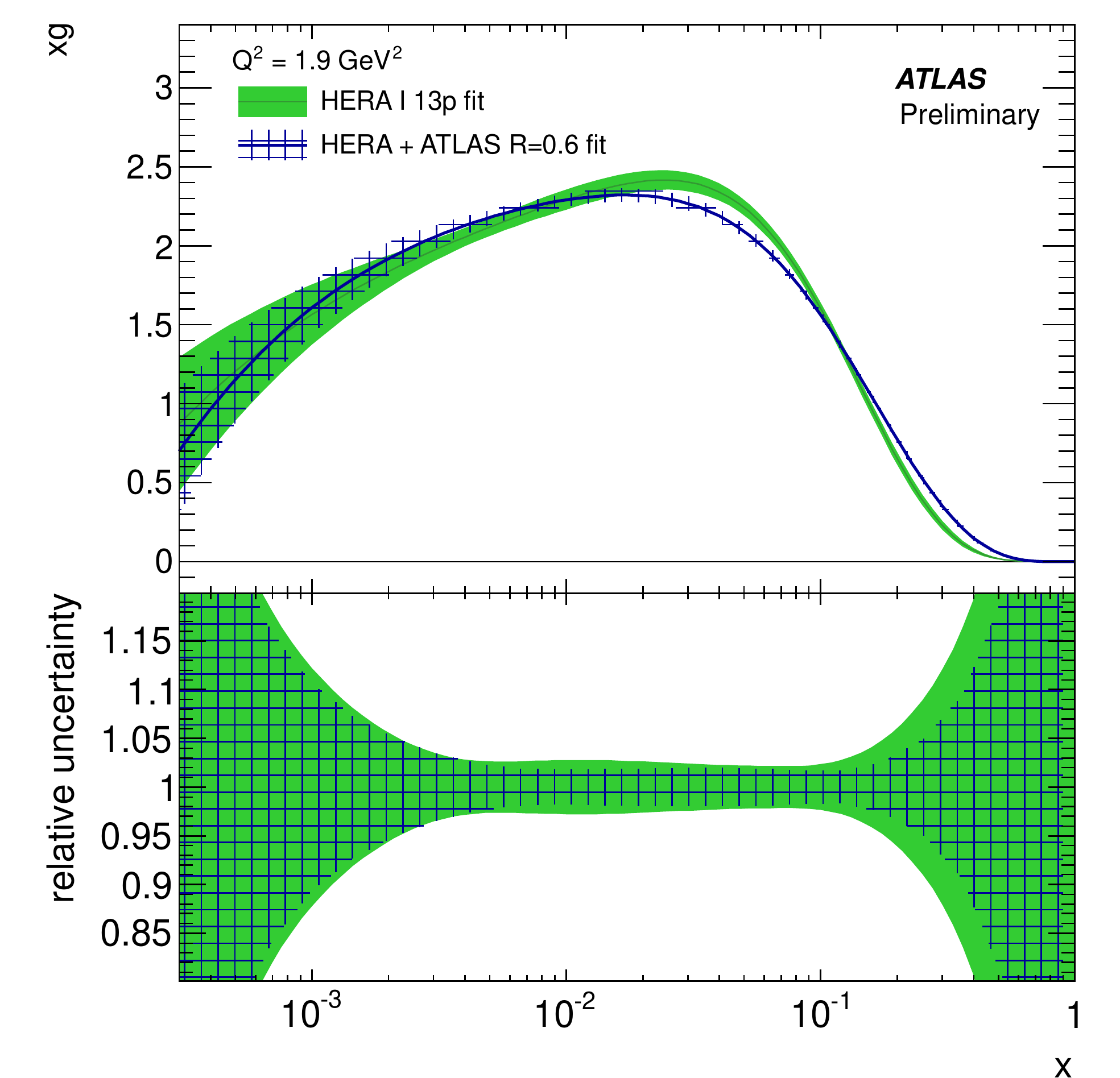}
\caption{Gluon density, shown together with its relative uncertainty and as a function of the fractional parton momentum,
  resulting from a fit to HERA data only (filled area) and when adding ATLAS jet data (hatched area) \cite{ATLAS-CONF-2012-128}.}
\label{fig:ATLAS-CONF-2012-128}
\end{figure}

Beginning of 2011, the LHC delivered proton-proton collisions at $\sqrt{s} =$~2.76~TeV,
which the ATLAS Collaboration used to measure the inclusive jet cross section up to jet $p_T$ of 430~GeV
\cite{ATLAS-CONF-2012-128}.
For given ranges in jet $p_T$ and rapidity, $y$,
these cross sections probe jet production at significantly different $x$ and $Q^2$
compared to $\sqrt{s} =$~7~TeV.
Despite the small size of the dataset (0.2~pb${}^{-1}$), the statistical uncertainty on these cross sections
at $\sqrt{s} =$~2.76~TeV is smaller than the total systematic uncertainty.
A crucial aspect here is that the jet reconstruction, energy scale and resolution are common to the analyses at 2.76
and 7~TeV, which means that the main systematic uncertainties can be assumed to be fully correlated.
This results in drastically reduced uncertainties on cross-sections ratios,
$\rho (p_T, y) = \sigma_{\text{jet}}^{\text{2.76 TeV}} (p_T, y) / \sigma_{\text{jet}}^{\text{7 TeV}} (p_T, y)$.
PDF uncertainties are found to dominate the uncertainty on the prediction for these ratios and to be generally larger than
the uncertainty on the measured ratio.
The ATLAS Collaboration studied the impact of the jet cross sections at $\sqrt{s} =$~2.76 and 7~TeV
with all their correlations using the HERAFitter framework.
Compared to parton distribution functions obtained based on HERA DIS data only,
the inclusion of the ATLAS jet data was found to result in a harder gluon distribution with a reduced uncertainty
at high $x$, which is illustrated in figure~\ref{fig:ATLAS-CONF-2012-128},
while slightly decreasing the high-$x$ sea-quark density.

A first $\alpha_S$ determination based on LHC data was performed using the inclusive jet cross section
from the ATLAS experiment at $\sqrt{s} =$~7~TeV up to jet $p_T$ of 600~GeV \cite{Malaescu:2012ts}.
The result obtained by
simultaneously fitting different rapidity bins up to $|y| <$ 4.4
is $\alpha_S (M_Z)$ = 0.1151 $\pm$ 0.0047 (stat.) $\pm$ 0.0014 ($p_T$ range) $\pm$ 0.0060 (jet size)
${}^{+0.0044}_{-0.0011}$ (scale) ${}^{+0.0024}_{-0.0018}$ (PDF) ${}^{+0.0009}_{-0.0034}$ (non-pert. corr.), where
the largest uncertainty arises from a disagreement seen between the values obtained with jet radii $R = 0.4$ and 0.6.

\begin{figure}
\centering
\includegraphics[width=.9\columnwidth,clip]{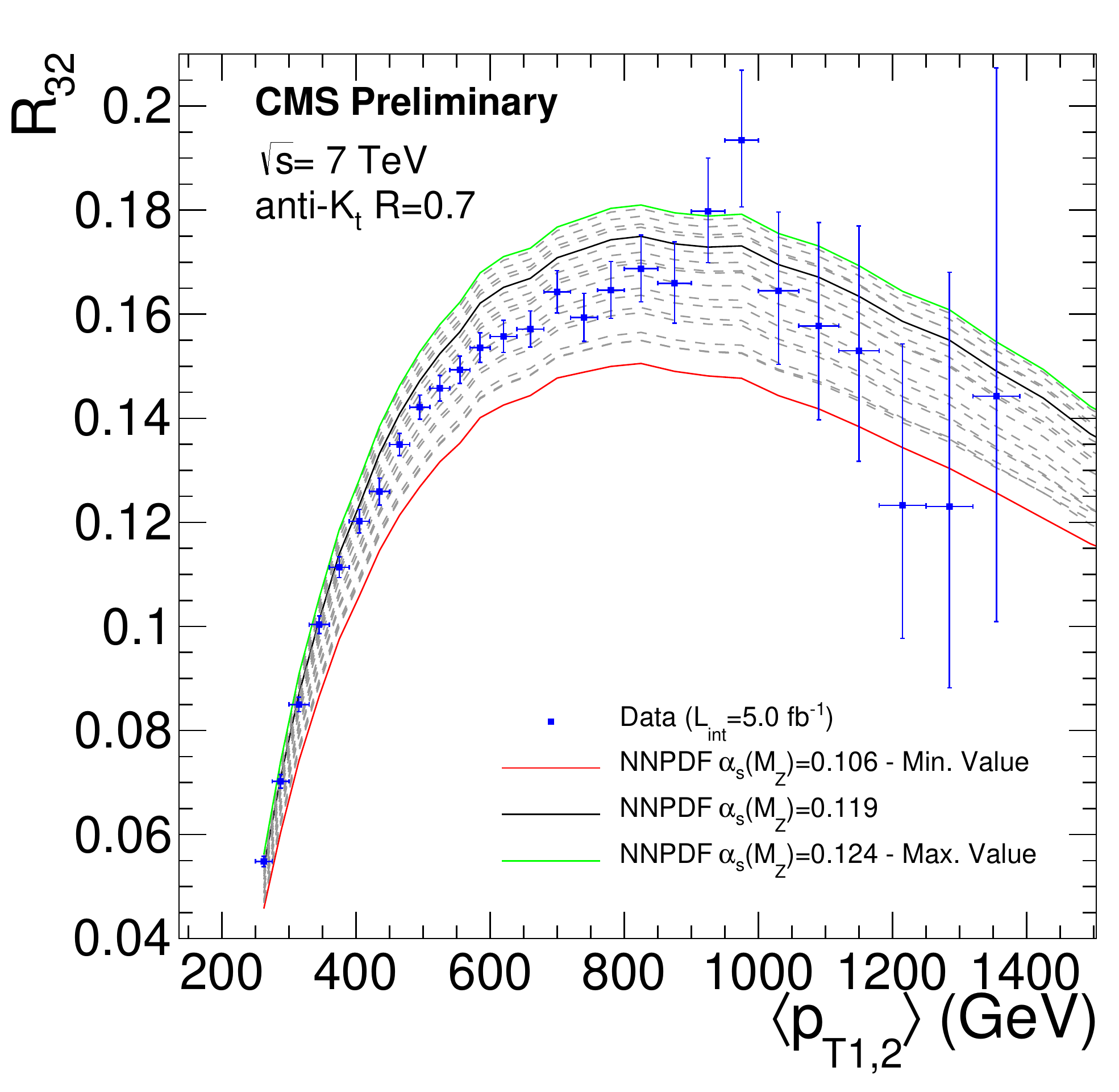}
\caption{Ratio of the inclusive 3-jet to 2-jet cross sections
  measured by the CMS Collaboration as a function of the average transverse momentum of the two leading
  jets and compared to the NLO prediction with different $\alpha_S (M_Z)$ values \cite{CMS-PAS-QCD-11-003}.}
\label{fig:QCD-11-003}
\end{figure}

The CMS Collaboration has recently performed a measurement of $R_{32}$, the ratio between the inclusive trijet and dijet
production rates \cite{CMS-PAS-QCD-11-003}.
Again, both the experimental and the theoretical uncertainties are drastically reduced by taking a cross-section ratio.
When selecting jets with $p_T >$ 150~GeV, the 2011 dataset allows probing the kinematic region
of 250 $< \langle p_{T1,2} \rangle <$ 1400 GeV, where $\langle p_{T1,2} \rangle$ is the average $p_T$ of the two
leading jets in the event.
Overall, the measured $R_{32}$ is well described by the NLO prediction.
Only when using the ABM11 PDF set, the prediction significantly underestimated the data up to $\langle p_{T1,2} \rangle$ of
600~GeV.
$R_{32}$ is proportional to $\alpha_S$, as illustrated in figure~\ref{fig:QCD-11-003},
and it has the advantage of a reduced dependence on assumption for the low-$Q^2$ evolution
made in the PDFs.
In order to reduce the impact of uncertainties related to choice and variation of the factorization and renormalization
scales, events with $\langle p_{T1,2} \rangle <$ 400~GeV are neglected when determining the best $\alpha_S$ in a
simultaneous $\chi^2$ fit to the different bins up to $\langle p_{T1,2} \rangle$ = 1.4 TeV,
which yields a logarithmic mean of 764~GeV for the considered range.
The result is then evolved downwards in energy to obtain
$\alpha_{S}(M_{Z})$ = 0.1143  $\pm$ 0.0064 (exp.)  $\pm$ 0.0019 (PDF)$\, _{-0.0000}^{+0.0050}$(scale). This result
is based on the NNPDF2.1 distribution functions.
Compatible results are obtained with CT10 and MSTW2008 PDF sets,
while the low gluon density in ABM11 would yield an $\alpha_{S}(M_{Z})$ larger than 0.1200, which is the upper edge
of the $\alpha_{S}(M_{Z})$ values supported by that PDF group.
\begin{figure}
\centering
\includegraphics[width=\columnwidth,clip]{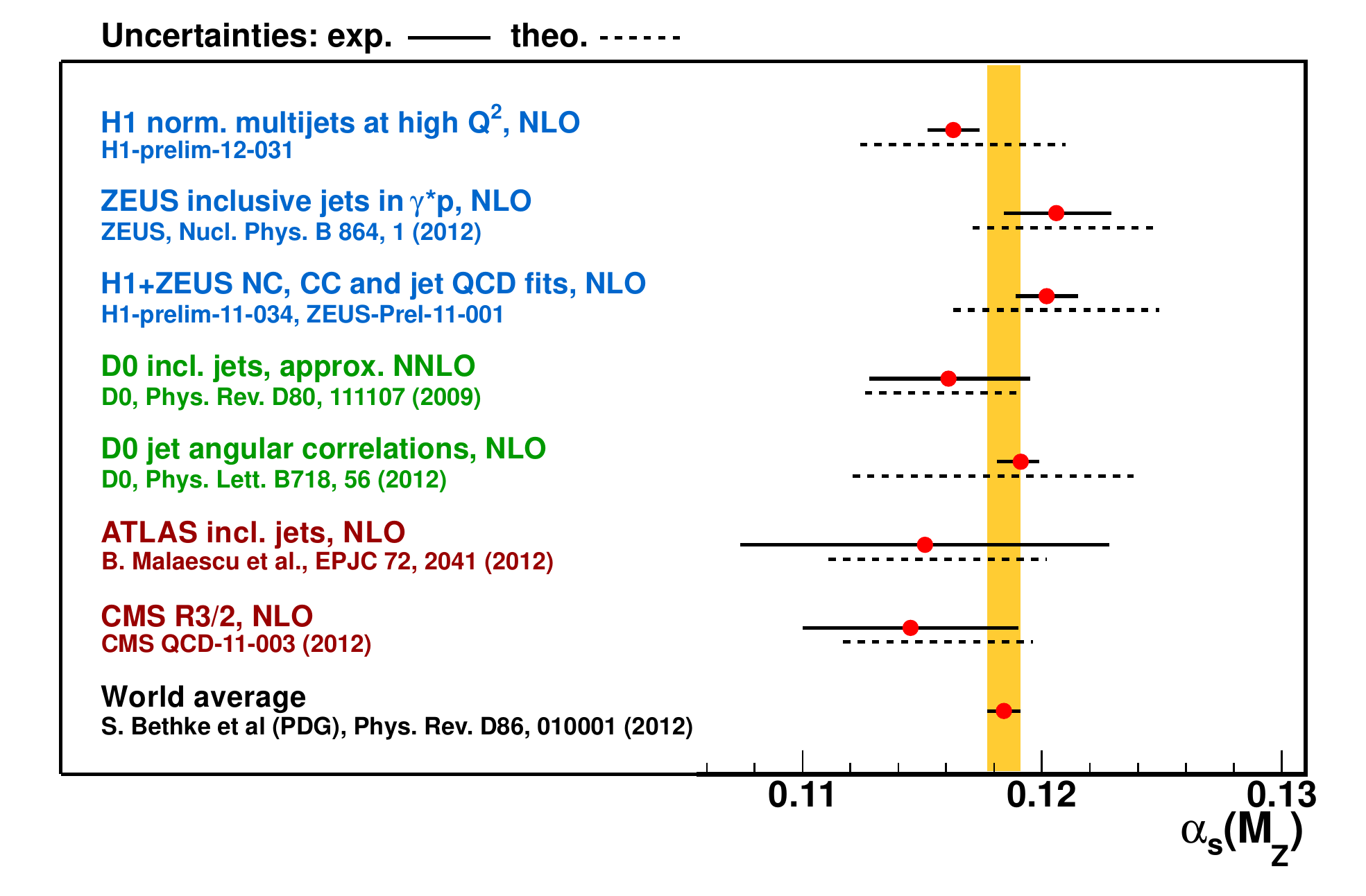}
\caption{All jet-data based $\alpha_S (M_Z)$ results discussed in this article compared to the latest world average
for $\alpha_S (M_Z)$. \cite{Kogler}}
\label{fig:alphas_summary}
\end{figure}

\bigskip

\section{Conclusions}
\label{sec:conclusions}
Cross sections for jet and multijet production have been measured with high precision at HERA, the Tevatron and the
LHC, probing QCD predictions over impressive ranges in $x$ and $Q^2$.
They provide strong constraints on the gluon PDF at medium to high $x$,
allow measuring $\alpha_S$ up to the TeV scale already and disentangling gluon PDF and $\alpha_S$.
Figure~\ref{fig:alphas_summary} compares recent jet-based $\alpha_S (M_Z)$ results,
all of them perfectly compatible with the latest world average but affected by significant theory uncertainties.
Given the sizable uncertainties related to choice and variation of factorization and renormalization scales as well as from
non-perturbative corrections, all such jet measurements and subsequent extractions of QCD parameters are eagerly
awaiting cross-section predictions beyond NLO.

\bibliography{references}

\end{document}